\documentclass[a4paper]{article}

\usepackage{fullpage,url,amsmath,amsfonts,amssymb,tedmath}
\newcommand{\ord}{\rm{OrderMod}}
\title{Entanglement-assisted quantum error-correcting codes from units}
\author{Ted Hurley\footnote{National University of Ireland
Galway. Ted.Hurley@NuiGalway.ie}, Donny Hurley\footnote{Institute of
Technology, Sligo. hurleyd@yahoo.com }, Barry
Hurley\footnote{barryj\_2000@yahoo.co.uk}} \date{}
\begin{document}
\maketitle
\begin{abstract}\let\thefootnote\relax 
 Entanglement-assisted quantum error-correcting codes
(EAQECCs) to desired rate, error-correcting capability and maximum
 shared entanglement are constructed. Thus for a required  rate $R$, required error-correcting  capability to correct $t$ errors, mds (maximum distance separable)
 EAQECCs of the form $[[n,r,d;c]]$ with $R=\frac{r}{n},  d\geq (2t+1), c = (n-r), d= (n-r+1)$ are constructed. Series of such codes may be constructed where the rate and the relative distance approach non-zero constants as $n$ approaches infinity. The codes may also be constructed over prime order fields in which { modular arithmetic} may be employed. 
\end{abstract}
\section{Introduction}

  Entanglement-assisted quantum error-correcting codes, EAQECCs, were
introduced by Brun, Develak and Hseih \cite{eaqecc5}  as an alternative
to the CSS constructions. Definitions, properties 
and background may be found in  book chapter form \cite{bookeaqecc}. 

The CSS construction and results of
\cite{hurley} are used in \cite{hurley101} to construct mds quantum
error-correcting codes to given specifications.  Efficient decoding algorithms are deduced from \cite{hurley}. 
The methods are adapted to construct 
EAQECCs to desired rate, error-correcting capability and maximum
 shared entanglement. For a required  rate $R$, required error-correcting  capability to correct $t$ errors, mds 
 EAQECCs of the form $[[n,r,d;c]]$ with $R=\frac{r}{n},  d\geq (2t+1), c = n-r, d= n-r+1$ are constructed. Methods of \cite{hurley} give efficient decoding algorithms. Many series of such codes may be constructed where the rate and the relative distance approach non-zero constants as $n$ approaches infinity. Such codes may be  constructed over prime order fields so that modular arithmetic may be employed.

 Codes with positive rate rate are constructed when $2r > n$ in which case {\em catalytic codes} may be deduced by \cite{eaqecc3}. 


Analogous to the CSS constructions, EAQECC constructions are developed  in
\cite{eaqecc2} and \cite{eaqecc1}; these are given as follows. 


\begin{construction}\label{construct1}\cite{eaqecc2} 
Let $\mathcal{C}$ and $\mathcal{D}$ be $[n,k_1,d_1]_q$ and
 $[n,k_2,d_2]_q$ linear codes with check matrices $H,K$
 respectively. Then an entanglement-assisted $[[n,k_1+k_2-n+c,
 \min\{d_1,d_2\};c]]_q$ quantum error-correcting code can be constructed
 where $c= \rank HK\T$.
\end{construction}


The suffix $q$, which means the codes are over the field $GF(q)$ of $q$
elements,  will be omitted from now on. 

The {\em rate} of an $[[n,r,d;c]]$ EAQECC is $\frac{r}{n}$ and the {\em
net rate} of the code is $\frac{r-c}{n}$. If the net rate is positive
then by \cite{eaqecc3} {\em catalytic codes} may be deduced. The methods here allow for the production of series of
EAQECCs with desired maximum possible positive net rate.

Good EAQECCs from circulant matrices are derived in \cite{eaqecc10}.


\section{Theorems and Constructions}

A Vandermonde (square) matrix is a matrix of the form

$$V_n= \begin{pmatrix} 1 & x_1 & x_1^2 & \ldots & x_1^{n-1} \\ 1& x_2 &
	x_2^2 & \ldots & x_2^{n-1} \\ \vdots & \vdots & \vdots & \vdots
	&\vdots \\ 1& x_n & x_n^2 & \ldots & x_n^{n-1}\end{pmatrix}$$

 The entries $x_i$
are usually taken to be elements in some field. The determinant of $V_n$ is $\prod_{i<j}(x_i-x_j)$. 

A particularly nice Vandermonde matrix is when the $x_i$ are the
different $n^{th}$ root of unity, that is when $x_i = \om^{i-1}$ where
$\om^n=1$ and $\om^j\neq 1$ for $1\leq j < n$. Not every field has an
$n^{th}$ root of unity.  Let $\om$ be a primitive $n^{th}$ root of unity
in a field $\F$; {\em primitive} as usual means that $\om^n=1$ but
$\om^r \neq 1$ for $1\leq r < n$. For such an $\om$ to exist in $\F$ it
is necessary that the characteristic of $\F$ does not divide $n$ and in
this case $n$ has an inverse in $\F$.  When the characteristic does not
divide $n$ the field may be extended to include an $n^{th}$ root of
unity.

The Fourier $n\ti n$ matrix, relative to $\om$, is the $n\ti n$ matrix

$$F_n= \begin{pmatrix} 1 &1 &1 & \ldots & 1 \\ 1 & \om & \om^2 & \ldots
       & \om^{n-1} \\ 1&\om^2 & \om^{2(2)} & \ldots & \om^{2({n-1})} \\
       \vdots & \vdots & \vdots & \ldots & \vdots \\ 1 & \om^{n-1} &
       \om^{(n-1)2} & \ldots & \om^{(n-1)(n-1)} \end{pmatrix} $$.

The inverse of $F_n$ is obtained directly by replacing $\om$ by
$\om^{n-1}$ in the above formula and dividing by $n$; $n^{-1}$ exists in
$\F$. The inverse of a general Vandermonde matrix is not so nice but a
formula exists. 

The rows, in order, of a Fourier $n\ti n$ matrix $F_n$ under consideration will
be denoted by $\{e_0, e_1, \ldots, e_{n-1}\}$. Thus $e_i = (1, \om^i,
\om^{i(2)}, \ldots, \om^{i(n-1)})$ for the primitive $n^{th}$ root of
unity $\om$ used to build $F_n$. Let $F_n^*$ denote the matrix obtained
by replacing $\om$ by $\om^{n-1}$. Then $F_nF_n^* = nI_{n\ti n}$. Denote
the columns in order of $F_n^*$ by $\{f_0,f_1,\ldots, f_{n-1}\}$. Then
the following fundamental equation is basic:

$$F_nF_n^* = \begin{pmatrix} e_0 \\ e_1 \\ \vdots \\e_{n-1}\end{pmatrix}
(f_0,f_1,\ldots, f_{n-1})= nI_{n\ti n}$$

In particular $e_if_j = n\de_{ij}$, where $\de_{ij}$ is the Kronecker
delta. Now it is easy to check that $f_i = e_{n-i}\T$ and
$e_i=f_{n-i}\T$, where $e_n=e_0, f_n=f_0$.  Thus $e_ie_{n-i}\T =
e_if_{i} = n$ and $e_ie_j\T = e_if_{n-j}= 0$ for $j\neq (n-i)$.

The following is implict in  \cite{hurley}: 
\begin{theorem}{\cite{hurley}}\label{seq2} Let $\mathcal{C}$ be a code generated by taking $r$ consecutive rows of a Vandermonde $n\ti n$ matrix. Then $\mathcal{C}$ is an mds (maximum distance separable) $[n,r,n-r+1]$ code.
  \end{theorem}

In particular we have:

\begin{theorem}{\cite{hurley}}\label{seq1} Let $\mathcal{C}$ be a code generated by taking $r$ consecutive rows of a Fourier $n\ti n$ matrix.  Then $\mathcal{C}$ is an mds $[n,r,n-r+1]$ code.
\end{theorem}
 For the Fourier matrix this can be extended as follows: 
\begin{theorem}{\cite{hurley}}\label{seq} Let $\mathcal{C}$ be a code generated by taking any $r$ rows  in arithmetic sequence with arithmetic difference $k$ satisfying $\gcd(n,k) = 1$ of a Fourier $n\ti n$ matrix. Then  $\mathcal{C}$ is an mds  $[n,r,n-r+1]$ code.
\end{theorem} 

Let $\mathcal{C} = <e_0,e_1, \ldots, e_{r-1}>$ be the code generated by
taking the first $r$ rows of a Fourier matrix $F_n$. 
Then $\mathcal{C}$ is an $[n,r,n-r+1]$ mds code by
Theorem \ref{seq1}. Let $\mathcal{D} = <e_{n-r+1}, e_{n-r+2}, \ldots,
e_{n-1}, e_0>$. Then also $\mathcal{D}$ is an $[n,r,d]$ mds code.

Let $H\T= (f_{n-1},f_{n-2}, \ldots, f_{r})= (e_{1}\T, e_{2}\T, \ldots,
e_{n-r}\T)$. Then $H= \begin{pmatrix} e_1 \\ e_2 \\ \vdots \\
e_{n-r}\end{pmatrix}$ is a check matrix for $\mathcal{C}$.

Let $K\T = (f_1, f_2, \ldots, f_{n-r})$ and then $K$ is a check matrix
			       for $\mathcal{D}$.

The following is then clear from $e_if_j = n\de_{ij}$.

\begin{proposition} $HK\T= nI_{n-r}$. 
\end{proposition}
\begin{corollary} $HK\T$ has rank $n-r$. 
\end{corollary}

\begin{theorem}
  By Construction \ref{construct1} \cite{eaqecc2} a maximum EAQECC
  $[[n,2r-n+c,d;c]]$ code is constructed from $\mathcal{C}$ and
  $\mathcal{D}$ with $d=n-r+1, c=n-r$.
\end{theorem}

Note that $c=n-r$ so the code is an $[[n,r,d;n-r]]$ code. The rate is
$\frac{r}{n}$ and the net rate is $\frac{2r-n}{n}$. Thus when $2r> n$,
codes with net rate greater than $0$ are produced and as shown in
\cite{eaqecc3} catalytic codes may then be formed.
\begin{corollary} When $2r > n$, catalytic codes may be constructed.
\end{corollary}

The `shared entanglement' consists essentially of those rows not used in the
construction of the code $\mathcal{C}$.

From methods in \cite{hurley} the codes have efficient decoding algorithms.


By taking $\mathcal{C}= <e_1,e_2,\ldots,e_{r}$ and $\mathcal{D}=
<e_{n-r},e_{n-r+1}, \ldots, e_{n-1})$, or more generally by taking
$\mathcal{C} = <e_i,e_{i+1}, \ldots,e_{i+r-1}>$ and $ \mathcal{D}=
<e_{n-(i+r-1)}, e_{n-i-r+2} \ldots, e_{n-(i-1)}>$, codes with check
matrices $H,K$ are derived so that $HK\T=nI_{n-r}$ from which it follows
that $\rank HK\T = n-r$. In these cases also mds EAQECCs $[[n,r,d;n-r]]$
may be formed by Construction \ref{construct1}.

\subsection{Summary of general construction}
\begin{enumerate} \item Form a Fourier or Vandermonde $n\ti n$ matrix.
\item Choose the first $r$ rows to form $\mathcal{C}$ and the last $r$ rows to form $\mathcal{D}$.
\item The check matrices $H,K$ for $\mathcal{C}, \mathcal{D}$ satisfy $\rank HK\T = n-r$.
\item Use EAQECC construction \ref{construct1} with $\mathcal{C}, \mathcal{D}$  to form an  $[[n,r,d;c]]$ mds EAQECC.
\item Required  rates and required error-correcting capability may be deduced by constructing an appropriate Fourier/Vandermonde matrix.
  \item $\mathcal{C}$ may be formed from any $r$ consecutive rows of the matrix and then $\mathcal{D}$ may be chosen as explained above so that the check matrices $H,K$  of $\mathcal{C},\mathcal{D}$ satisfy $\rank (HK\T) =n-r$. 
\end{enumerate}

For a given rate, given
error-correcting capability and shared entanglement, mds EAQECCs are   
constructed.

\subsection{Any shared entanglement} The full maximum possible shared entanglement may not be required. It is easy to modify the above to obtain codes of the form $[[n,2r-n+c,d;c]]$ 
for $1\leq c
\leq n-r$ by choosing $\mathcal{C}$ as before and then finding the
appropriate rows of the matrix for constructing $\mathcal{D}$ so that $\rank HK\T = c$, for $1\leq c \leq n-r$,  
where $H$ is a check matrix for $\mathcal{C}$ and $K$ is a check matrix
for $\mathcal{D}$. The details are omitted; see \ref{trist} below for
prototype example. 
These codes are also mds EAQECC codes as 
$n-(2r-n+c) + c = 2(d-1) $ exactly. The net rate is $\frac{2r-n}{n}$
which is positive for $2r> n$. Shared  entanglement as required is available.

\subsection{Prototype Examples}
The following are illustrative and prototype examples of the general
constructions. 
\begin{enumerate}

  \item\label{trist1} Suppose a $[[n,r,d;c]]$ code is required with $\frac{r}{n}=R=
		      \frac{7}{8}$, error-correcting capability such
		      that $d\geq 11$ and (maximum) entanglement $c=
		      n-r$.  Then it is required that
	$n=\frac{d-1}{1-R}$ which requires $n=80$ for
		      $d=11$. Thus a $[[80,70,11;10]]$ code is required 
		      where $10=80-70$ is the size of the check
		      matrix. Let $F_{80}$ be a $80\ti 80 $ Fourier or
		      Vandermonde matrix over a suitable field. Let the
		      rows of $F_{80}$ in order be denoted by
		      $\{e_0,e_1, \ldots, e_{79}\}$.  Then take as
		      $\mathcal{C}$ the code generated by the first $70$
		      rows of $F_{80}$ and for $\mathcal{D}$ the code
		      generated by the `last' $70$ rows of $F_{80}$,
		      that is $\mathcal{C}= <e_0, e_1, \ldots, e_{69}>$
		      and $\mathcal{D} = <e_{11}, e_{12}, \ldots,
		      e_{79}, e_{80}>$ where $e_{80}=e_0$. The check
		      matrices $H,K$ for $\mathcal{C}, \mathcal{D}$
		      satisfy $HK\T=80 I_{10}$ and thus $\rank HK\T =
		      10$. By Construction \ref{construct1} an mds 
		      EAQECC $[[80,70,11;10]]$ is constructed. The rate
		      is $\frac{7}{8}$ and net rate is $
		      \frac{6}{8}$. Since the net rate is positive,   catalytic QECCs may be deduced by
		      \cite{eaqecc3}.

 Consider $GF(3^4)$. Now $3^4-1 = 80$ and so a Fourier $80\ti 80$ matrix
 $F_{80}$ exists over $GF(3^4)$; $F_{80}$ is formed from a primitive
 $80^{th}$ root of unity in $GF(81)$.

\item Suppose as in example \ref{trist1} above a $[[n,r,d;c]]$ code with
   rate $\frac{7}{8}$,  error-correcting capability $d$ such that
   $d\geq 11$ and shared entanglement $c=n-r$ are required
   but now insist the Fourier matrix has to be  over a
  prime field. As before get $n=\frac{d-1}{1-R}$. Choose the smallest
  $d$ such that $d\geq 11$ and $n+1$ is prime. The first such $d$ is
  $d=12$ in which case $n=88$ and $n+1=89$ with $89$ a prime. Then get
  $[[88,77,12;11]]$ code over $GF(89)=\Z_{89}$ and  the arithmetic
  is {\em modular arithmetic}, which is very efficient. Now \ord$(3,89)=88$ and so $3 \mod 89$ may be
  used as the primitive element in forming the Fourier $88\ti 88$ matrix
  over $GF(89)=\Z_{89}$. Form the Fourier $88 \ti 88$ matrix over $\Z_{89} = GF(89)$ with $3 mod 88$ as the primitive element. Form for example $\mathcal{C} = <e_0,e_1, \ldots, e_{76} >, \mathcal{D}= e_{12}, e_{13}, \ldots, e_{88}>$ (where $e_{88} = e_0$). The check matrices $H,K$ of $\mathcal{C},\mathcal{D}$ satisfy $\rank HK\T= 11$.  Use the EAQECC construction \ref{construct1} on $\mathcal{C},\mathcal{D}$
  to get the $[[88,77,12;11]]$ mds EAQECC. 

\item\label{trist} Suppose a length ${11}$ mds EAQECC is required
     with shared entanglement $c=1$ and can correct $2$ errors. Consider a Fourier $11\ti 11$
     matrix $F_{11}$ with rows $\{e_0,e_1, \ldots, e_{10}\}$. Let columns of
     $F_{11}^*$ be denoted by $f_0,f_1, \ldots f_{10}$. Let $\mathcal{C}$
     be the mds $[11,7,5]$ code generated by $\{e_0, e_1, \ldots, e_6\}$
     and $\mathcal{D}$ the $[11,7,5]$ code generated by $\{e_2, e_3,
     \ldots, e_8\}$. A check matrix for $\mathcal{C}$ is
     $H= \begin{pmatrix}e_1 \\ e_2 \\ e_3 \\ e_4\end{pmatrix}$ and the transpose of a
	  check matrix $K$ for $\mathcal{D}$ is
	  $K\T=(f_9,f_{10},f_0,f_1)$. Then $HK\T$ is a matrix with one
	  non-zero entry and so has rank $1$. Thus by \ref{construct1} a
	  $[[11,4, 5;1]]$ EAQECC is
	  constructed. The rate is $\frac{4}{11}$ and the net rate is
	  $\frac{3}{11}$. As the net rate is positive, catalytic codes
	  may be constucted.  The code as also an mds EAQECC as the
	  formula $n-r+c =
	  2(d-1)$ in this case is $11-4+1 = 8$ which is true. 

There exist  $11 \ti 11$ Fourier matrices over (i) $GF(2^{10})$,  
     (ii) $GF(3^5)$, (iii) $GF(5^5)$. Now $GF(3^5)$ is the smallest.

To get a $c=2$ EAQECC, take $\mathcal{C}$ as above and $\mathcal{D}=
     <e_3, e_4, \ldots, e_9>$. This then gives an EAQECC $[[11,5,5;2]]$
     code with net rate $\frac{3}{11}$.

\item Suppose a rate $R= \frac{4}{7}$ is required with capability of
      correcting $5$ errors and with maximum possible entanglement. Thus
      a $[[n,r,d \geq 11; c]]$ code is required with $R= \frac{4}{7}=
      \frac{r}{n}$ and $c = n-r$. Then get $n=\frac{d-1}{1-R}=
      \frac{7(d-1)}{3}$. Thus it is necessary that $3/(d-1)$ for $d\geq
      11$. Values for $d$ are $13,16, 19,...$. Now for $d=13$ this gives
      $n=28, r= 13$  and get a $[[28, 16,13;12]]$ code. This can be done
      over $GF(29)$ as $n+1=29$ is prime. Now the order of $2 \mod 29$
      is $28$ so $2 \mod 29$ may be used as the primitive element in the
      Fourier $28\ti 28$ matrix over $GF(29)$ from which the code may be
      constructed; the arithmetic is modular arithmetic in $\Z_{29} =
      GF(29)$. The rate is $\frac{4}{7}$ and the net rate is
      $\frac{16-12}{28}= \frac{1}{7}$. 
\end{enumerate} 


\paragraph{Comment:} How EAQECCs perform compared to QECC has been
      discussed in other publications, see for example \cite{eaqecc1}
      and others. 
Dual-containing codes with high error-correcting capacity are no longer
restricted, by constructions in \cite{hurley,hurley101}, and thus the comparisons are now
at a different level. Can entanglement in addition to required rate and
required error-correcting capability improve the performance? 

\end{document}